\documentstyle[times,pramana,epsf,floats]{ias}
\def\f{\phi^{--}}
\def\h{\phi^{\pm\pm}}
\begin{document}
\mark{{Associated single photons and doubly charged scalar}{S K Rai and B Mukhopadhyaya}}
\title{Associated single photons and doubly charged scalar at linear
$e^-e^-$ colliders}
\author{Biswarup Mukhopadhyaya and Santosh Kumar Rai
\footnote{Talk presented at LCWS06, Bangalore, India }}
\address{Harish-Chandra Research Institute, Chhatnag Road, Jhunsi,
Allahabad 211019, India}
\keywords{Doubly Charged Higgs, associated photon, ISR}
\pacs{}
\abstract{
Doubly charged scalars, predicted in many models having
exotic Higgs representations, can in general have lepton-number 
violating (LFV) couplings. We show that by using an associated 
monoenergetic final state photon seen at a future linear $e^- e^-$ 
collider, we can have a clear and distinct signature for a 
doubly-charged resonance. The strength of the $\Delta L=2$ coupling 
can also be probed quite effectively as a function of the recoil mass 
of the doubly-charged scalar.}
\maketitle

\vspace*{0.2in}
Doubly charged scalars arise in a number of physics scenarios 
\cite{THM,LRSM},
the most common models to accommodate such scalars are those with triplet 
Higgs. An added feature often associated with doubly-charged Higgs is 
the possibility of lepton-number violation. This basically consists in 
$\Delta L =2$ couplings with leptons of the form

\begin{equation}
 \sl{L}_Y = i h_{ij}\Psi^T_{iL}C\tau_2\Phi\Psi_{jL} + h.c.
\end{equation}
where $i,j=e,\mu,\tau$ are generation indices, the $\Psi$'s are the 
two-component left-handed lepton fields, and $\Phi$ is the triplet with 
$Y=2$ weak hypercharge. This leads to mass terms for neutrinos once the 
neutral component $\phi^0$ of $\Phi$ acquires a vacuum expectation 
value (vev):

\begin{equation}
\sl{M}^\nu_{ij}  \sim h_{ij} v'
\end{equation}

\noindent
$v'$ being the triplet vev.
Constraints on the $\rho$-parameter puts strong limits
on the the triplet vev \cite{LEP} translating into 
limits on the L-violation Yukawa couplings from the expected ranges of
neutrino masses. Such limits usually constrain the collider signals for 
doubly-charged scalars sought through $\Delta L = 2$ interactions.  

We point out the usefulness of looking for doubly-charged
scalars  in an $e^- e^-$ collider, in the radiative production channel. 
Resonant production of the $\f$ requires one to know
its mass with reasonable accuracy to start with, and tune the center-of-mass
energy of the colliding electrons accordingly. In addition, precise 
identification of a doubly-charged resonance will also depend on its
decay products, which depend on the parameters of the L-violating sector. 
In general, one can have the decays
\begin{itemize} 
\item $\f \to W^{-}W^{-},l^{-}l^{-},W^{-}\phi^{-},\phi^{-}\phi^{-}$
\end{itemize}

A degeneracy among the triplet components is often a consequence of 
theories, albeit in a model-dependent fashion. If we thus neglect the 
last two channels listed above, we still have 
the $W^{-}W^{-}$ and $l^{-}l^{-}$ channels, of which the first is controlled
by the triplet vev $v'$ and the second, by the coupling $h_{ll}$. When
the first mode is dominant, it requires careful analysis of the W-decay
products in order to isolate signatures of resonant production.
It is thus desirable to have supplementary channels in mind while looking
for doubly-charged scalars \cite{alan}. 
With this in view, we have calculated the rates for the process
$$
e^{-} e^{-} \longrightarrow \f \gamma \longrightarrow X \gamma
$$
\noindent
at a $\sqrt{s}=1$ TeV $e^-e^-$ machine, concentrating on the hard 
single photon in the final state. This photon
will be monochromatic if a doubly-charged resonance is produced, 
irrespective of what it decays into. Furthermore, one is no more required 
to tune the electron-electron center-of-mass energy at a fixed value. 
For our analysis, taking the radiative production of the scalar $\f$ 
as the benchmark process, we concentrate only on the flavor diagonal 
coupling $h_{ee}$. In our numerical estimate, we have chosen the 
coupling strength to be $h_{ee}=0.1$ which  respects the 
most stringent bounds coming from muonium-antimuonium conversion 
results which for flavor diagonal coupling is 
$h < 0.44~M_\h~{\rm TeV}^{-1}$ at $90\%$ C.L.  
The on-shell radiative production of a doubly-charged scalar gives an 
almost monochromatic photon of energy 
\begin{equation} 
E_\gamma = \frac{s - M_{\f}^2}{2\sqrt{s}}
\label{egamma}
\end{equation}
which stands out against the continuum background of the standard model
(SM). We assume the $h_{ii}$ couplings are of equal strength. 
The total decay width 
of $\f$ thus obtained is very miniscule ($\sim 1.2$ GeV) for a 1 TeV 
scalar mass when compared to the machine energy and allows one to use the 
narrow-width approximation.

\begin{figure}[htbp]
\centerline{
\epsfxsize=5.5cm\epsfysize=5.5cm\epsfbox{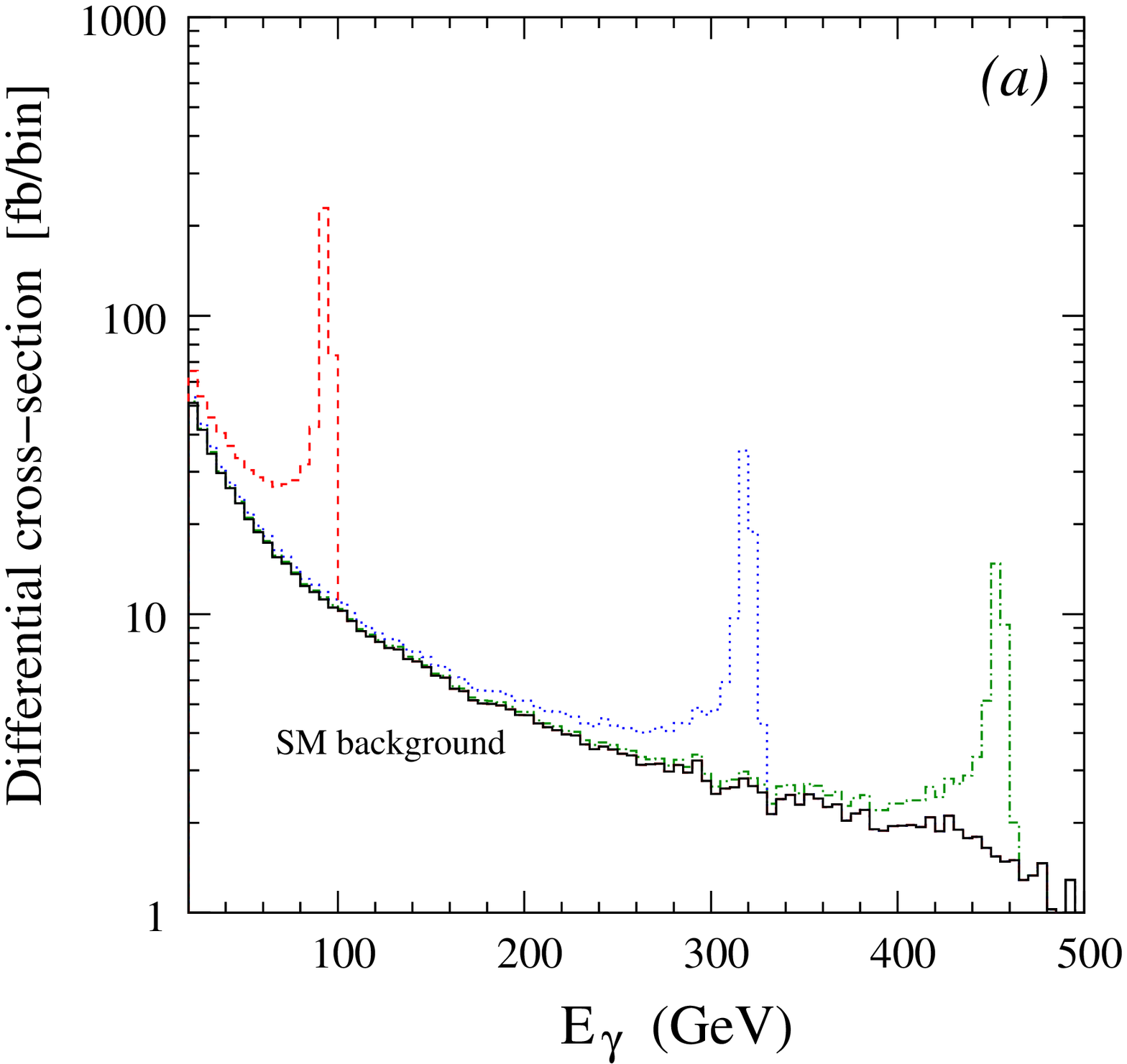}\epsfxsize=5.5cm\epsfysize=5.5cm\epsfbox{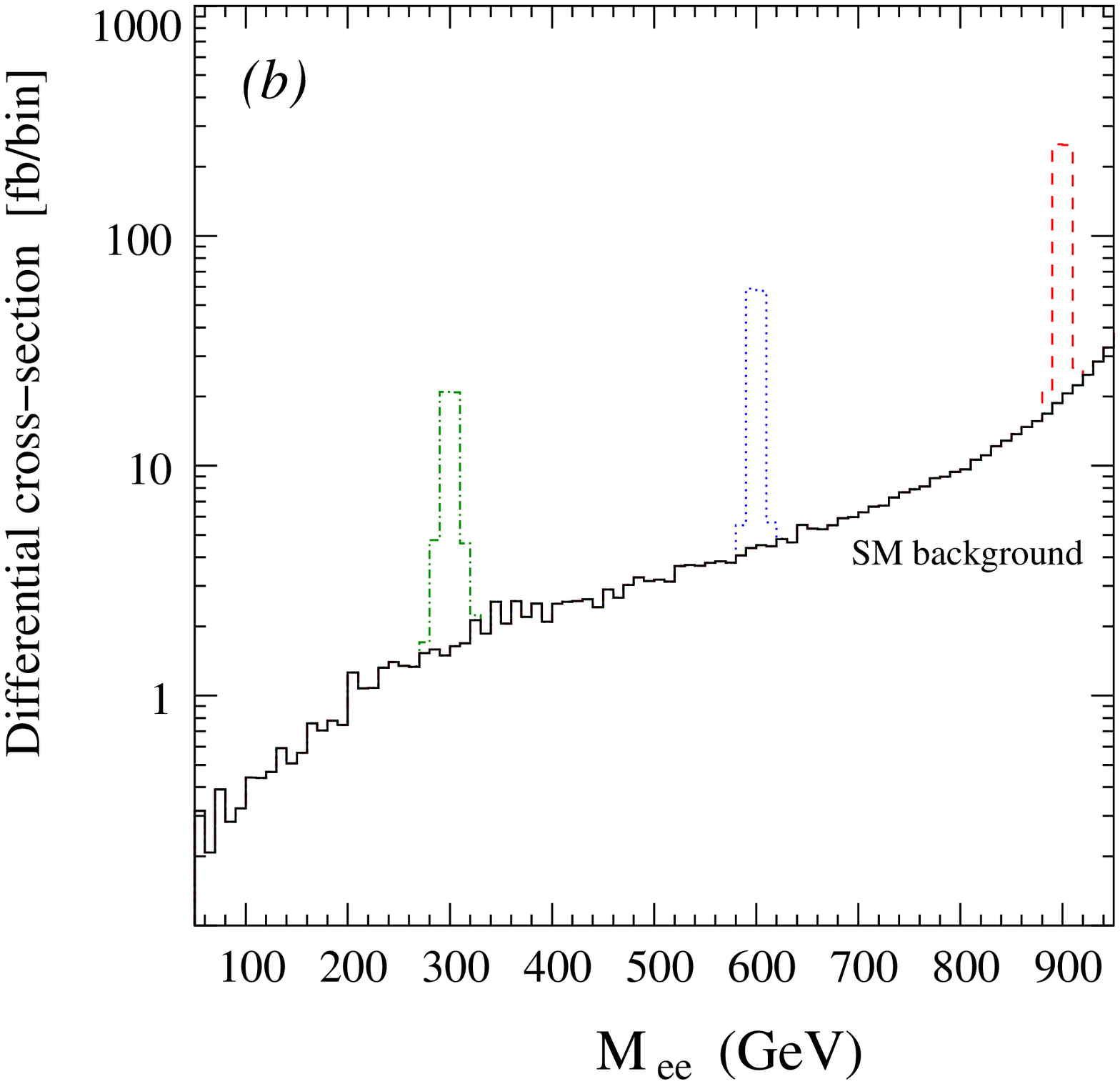}}
\caption{\sl\small Differential cross-sections against (a) photon energy 
$E_\gamma$ and (b) invariant mass of electron pair $M_{ee}$. 
The dash-dot-dash (green) line corresponds to $M_{\phi^{--}} = 300$ GeV, 
dotted (blue) line corresponds to $M_{\phi^{--}} = 600$ GeV and the 
dashed (red) line corresponds to $M_{\phi^{--}} = 900$ GeV
respectively. The binsize is chosen to be 5 GeV in (a) and 10 GeV in (b).}
\label{resonance1}
\end{figure}

The major SM background that contributes to the above process is the
radiative Moller scattering process: $e^- + e^- \to \gamma + e^- + e^-$ 
which, although a continuum background, could {\it prima facie} be
large enough to wash away the monochromatic peak. The event selection
criteria, therefore, are largely aimed at suppressing this continuum
background. We impose the following set of cuts. 
\begin{itemize}
\item Rapidity cut on the final state particles:~
$|\eta(e^-)| < 1.5 ~~~{\rm and}~~~|\eta(\gamma)| < 2.5$
\item minimum cut on energies:
$E(\gamma) > 20 ~{\rm GeV}$, $ E(e^-) > 5 ~{\rm GeV}$
\item to ensure that the final state particles are well separated in
space for the detectors to resolve events : $\delta R > 0.2$
\end{itemize}
Using the above cuts we make an estimate of the SM background and the 
signal. 
We focus on the main trigger, {\it viz.} the photon. In Fig
\ref{resonance1}(a) we show the distribution of the photon energy, 
where we have
superposed the differential cross-section for {\it signal+background} 
in each bin over the SM background. A pronounced peak can be seen in
the photon energy distribution, due to the monochromaticity of
the photon, corresponding to the recoil energy
against the scalar resonance through the relation of Eq.\ref{egamma}.
To make our analysis realistic, we have smeared the photon energy by a
Gaussian function whose half-width is guided by the resolution of the
electromagnetic calorimeter \cite{smear1,smear2} and also incorporated 
the effects of ISR which often results in substantial broadening of the 
peak. We show the resulting peak for three choices of scalar mass 
(300, 600, 900 GeV). Alternatively, in Fig \ref{resonance1}(b), we also
show the invariant mass distribution of the $ee$ pair for the above 
choice of parameters and as expected the distribution peaks corresponding 
to the mass of scalar. 

\begin{figure}[htbp]
\centerline{
\epsfxsize=5.5cm\epsfysize=5.5cm\epsfbox{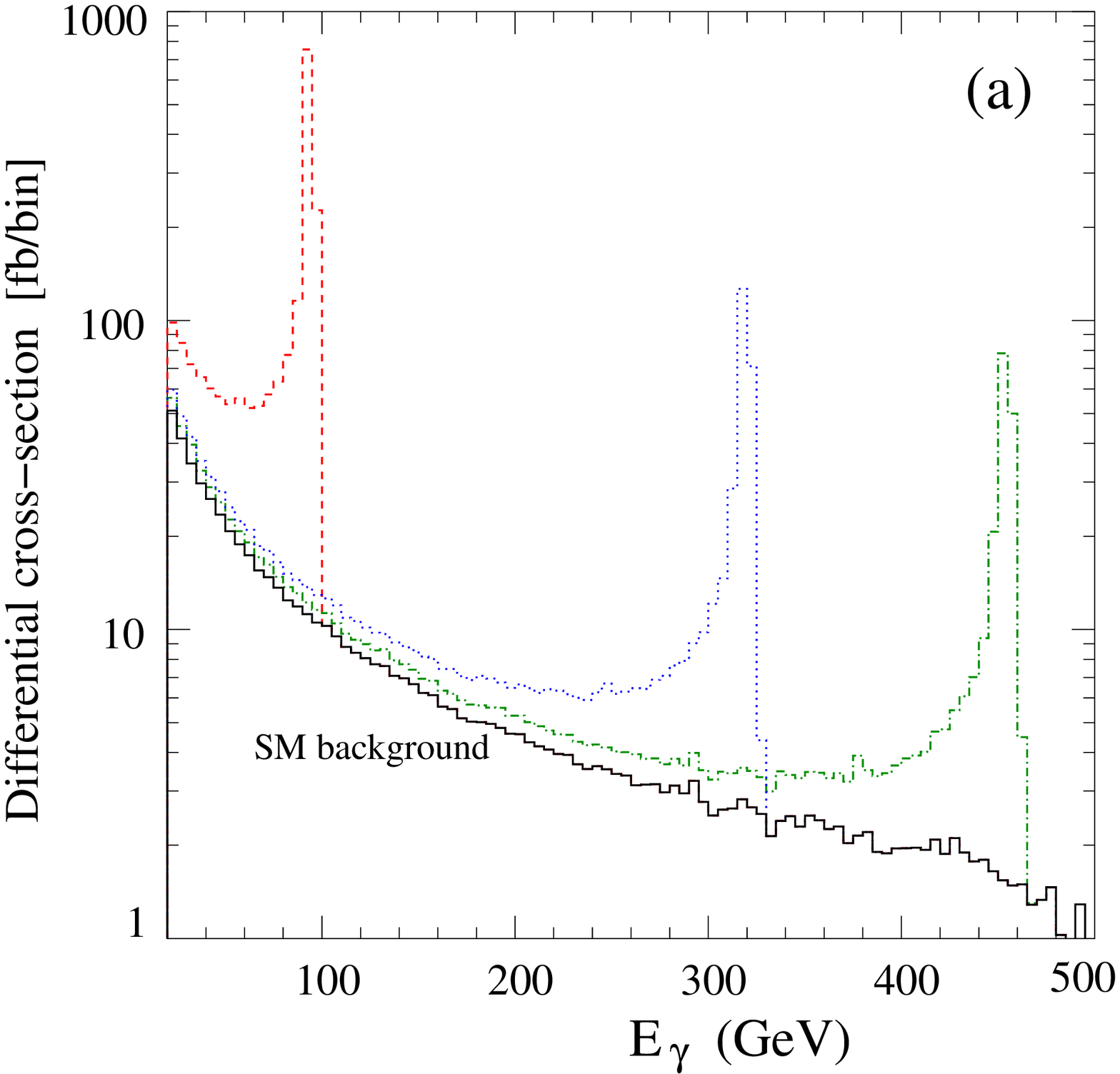}\epsfxsize=5.5cm\epsfysize=5.5cm\epsfbox{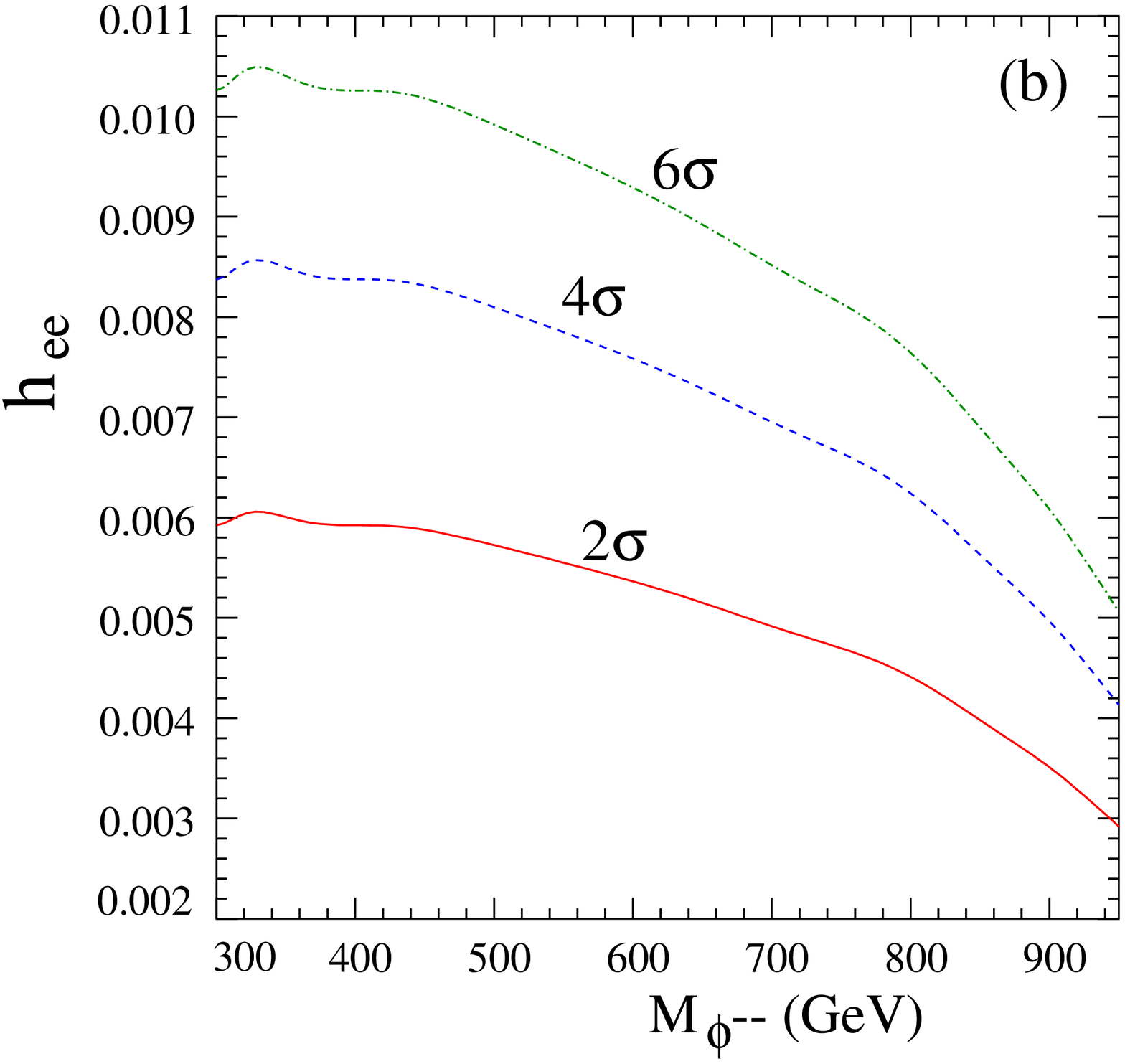}}
\caption{\sl\small {\bf (a)} Differential cross-sections against 
photon energy $E_\gamma$ when $\phi^{--} \to X$(anything). The 
dash-dot-dash (green) line corresponds to $M_{\phi^{--}} = 300$ GeV, 
dotted (blue) line corresponds to $M_{\phi^{--}} = 600$ GeV and the 
dashed (red) line corresponds to $M_{\phi^{--}} = 900$ GeV respectively.
{\bf (b)} Illustrating the reach of the coupling constant at which the 
resonances in the $E_\gamma$ distribution can be identified over the
fluctuations in the SM background. The assumed luminosity is 
100 fb$^{-1}$.}
\label{resonance2}
\end{figure}

In Fig \ref{resonance2}(a) we plot the energy distribution of the 
photon once again but here we only look at the final
state hard transverse photon in $e^- e^- \to \gamma + \f \to \gamma + X$
The distribution again shows peaks corresponding to the
recoil against the massive scalars, irrespective of the knowledge of the
decay products of the scalar. In fact our signal here receives a
relative boost as it 
is not suppressed by considering any further decay since the 
$BR(\phi^{--}\to X)= 100\%$. The fact that looking at a single 
photon against the backdrop of a continuum background makes it possible to
identify a LFV ($\Delta L=2$) process in a model independent way, makes
this signal worth studying at a future $e^- e^-$ collider and running
the linear collider in this mode. 

Since the rates for the signal depend directly on the $eeH$ coupling 
squared, in Fig \ref{resonance2}(b) we show the strength of the coupling 
for which the peaks would stand out against the fluctuations in the SM 
background. In our analysis we have assumed a luminosity of 
${\mathcal L} = 100~{\rm fb}^{-1}$. The fact that we are not looking at
any specific final state arising from $\f$ decay improves the reach of
this search channel. However, if a direct resonance is excited then that
would invariably translate into a much stronger probe of the coupling
strength. Nonetheless, our analysis is not dependent on
the tuning of the $\sqrt{s}$ of the machine to hit a resonance 
and hence serves as a more robust proposition. For luminosity higher
than what we have used, this reach can be further enhanced.

To summarise, the cleanliness of central photon detection at a high energy
linear collider can be very helpful in identifying a doubly-charged
scalar.  The peaks in the hard photon energy can
be helpful in two ways. First, one does not need to tune the two
electron beams, and can therefore work without a prior knowledge of the
$\f$ mass. Secondly, this method is shown to work even if the $\f$
dominantly decays into states that are not clean enough for the
resonance to be identified. Thus, as soon as one succeeds in reducing
the SM backgrounds, one can clearly see $\Delta L=2$ interactions, 
just by looking at the
accompanying hard photon. Not only doubly-charged scalars but also more
exotic resonances such as bileptons are amenable to detection in
this manner.

This talk was based on on earlier work and for further details one is
suggested to look at reference \cite{SKR}.

\end{document}